\documentclass[11pt]{article}
\usepackage{moriond,epsfig}
\usepackage{graphicx}
\usepackage[dvips]{color}
\usepackage{latexsym}
\usepackage{wasysym}

\def\be{\begin{equation}}
\def\ee{\end{equation}}
\def\bea{\begin{eqnarray}}
\def\eea{\end{eqnarray}}

\newcommand{\white}[1]{\textcolor[gray]{1.0}{#1}}
\newcommand{\comment}[1]{}

\begin{document}
\vspace*{4cm}
\title{CHARM PRODUCTION IN EP-COLLISIONS}

\author{ J. Wagner, representing the ZEUS and H1 Collaborations}

\address{DESY FH1, Notkestrasse 85\\ D-22603 Hamburg, Germany}

\maketitle\abstracts{
Recent results on the production of open charm in electron proton scattering at HERA
are reviewed. Results on the fragmentation of charm are presented and compared to the measurements
at $e^+e^-$ colliders. Recent results on the charm contribution to the proton structure function 
$F_2$ are shown. Furthermore, measurements of inclusive $D^*$ meson, $D^*$ + jet and $D^*$ + muon 
production are presented. These results are compared with perturbative QCD 
calculations at next-to-leading order as well as with Monte Carlo predictions.
}

\section{Introduction}
At HERA, electrons (or positrons) of energy 27.5 GeV are collided with 920 GeV protons, 
providing a center-of-mass energy of $\sqrt{s}\approx 318\;\mbox{GeV}$. 
During the HERA-I run, the experiments H1 and ZEUS each
collected data corresponding to an integrated luminosity of about $110\;\mbox{pb}^{-1}$, thus allowing
significant tests of perturbative QCD for heavy quark production.\\
In electron proton collisions, heavy quarks are predominantly produced via the photon gluon fusion 
(PGF) mechanism, in which a photon emitted by the incoming electron interacts with a gluon in the proton 
forming a quark anti-quark pair. The cross section is largest in photoproduction, in which the
exchanged photon is almost real $Q^{2}\sim 0$, and decreases towards the region of deep inelastic
scattering (DIS) which is at larger values of $Q^2$ ($Q^2\apprge 2\,\mbox{GeV}^2$).\\
Heavy quark production can be described by factorising the process into three parts: 
the proton structure, the hard partonic interaction and the fragmentation of the final state quarks
into hadrons. The structure function $F_2^{c\bar{c}}$ denotes the contribution from events with charm
to the inclusive proton structure function $F_2$. The hard interaction process is calculable
in perturbative QCD as the heavy quark mass provides a hard
scale.\\
The HERA data are compared with next-to-leading order (NLO) calculations in the massive
(see \cite{gpNLO1} for photoproduction and \cite{DISNLO} for DIS), massless \cite{gpNLO2} and
matched \cite{gpNLO3} schemes, which are expected to hold in different limits of the scale 
$\mu$ used for the calculation of the hard cross section.
In addition the data are compared with the Monte Carlo generators PYTHIA\cite{Pythia}, 
which implements the DGLAP evolution equation \cite{DGLAP} and CASCADE\cite{Cascade} which is based on 
the CCFM evolution equation\cite{CCFM}.

\section{Fragmentation}
\begin{figure}[tb]
\setlength{\unitlength}{1cm}
\begin{center}
\begin{picture}(16,9.5)
\put(-1.2,-0.6){\epsfig{file=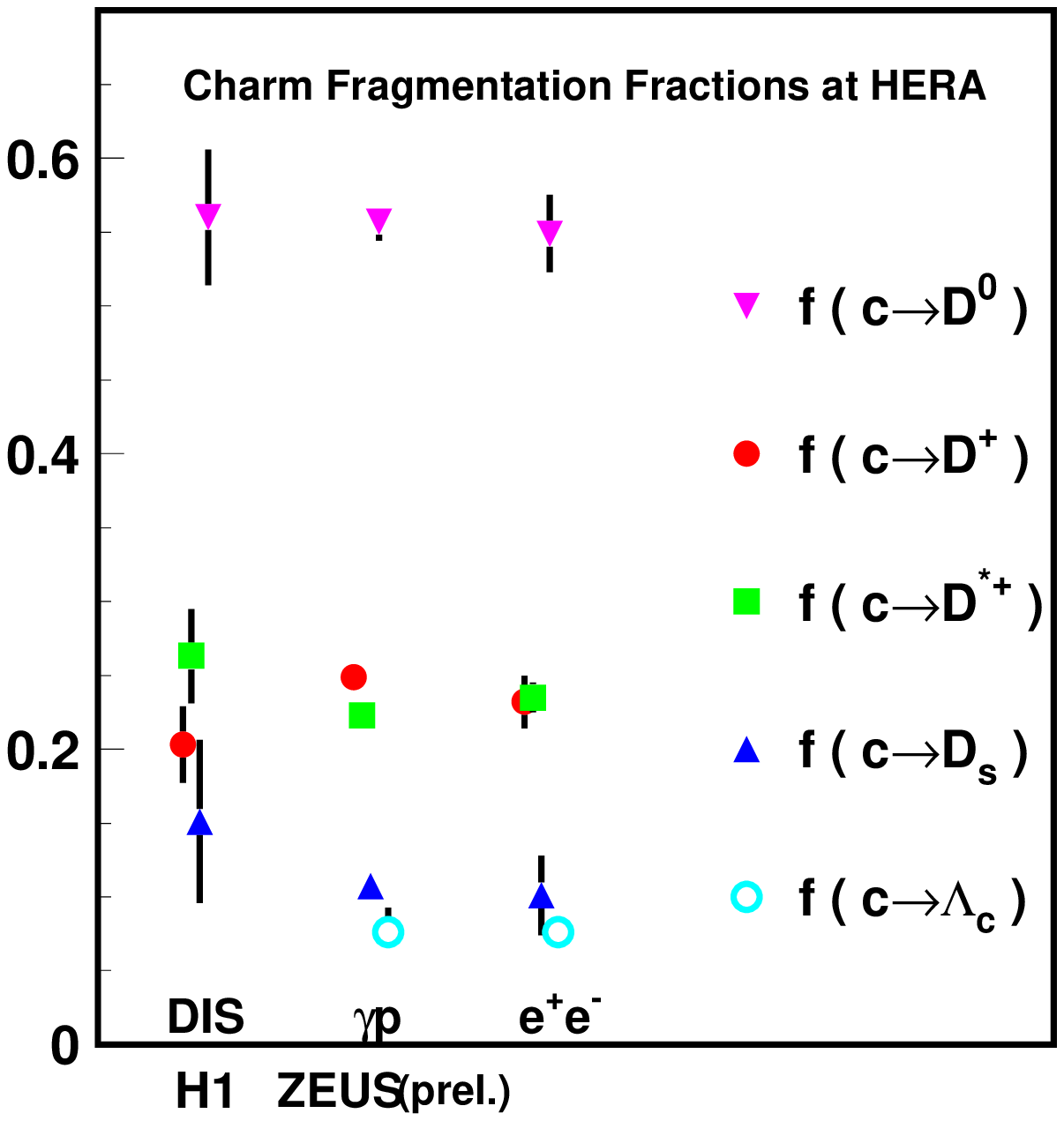, width=9.3cm,clip=true}}
\put(8.3,-0.3){\epsfig{file=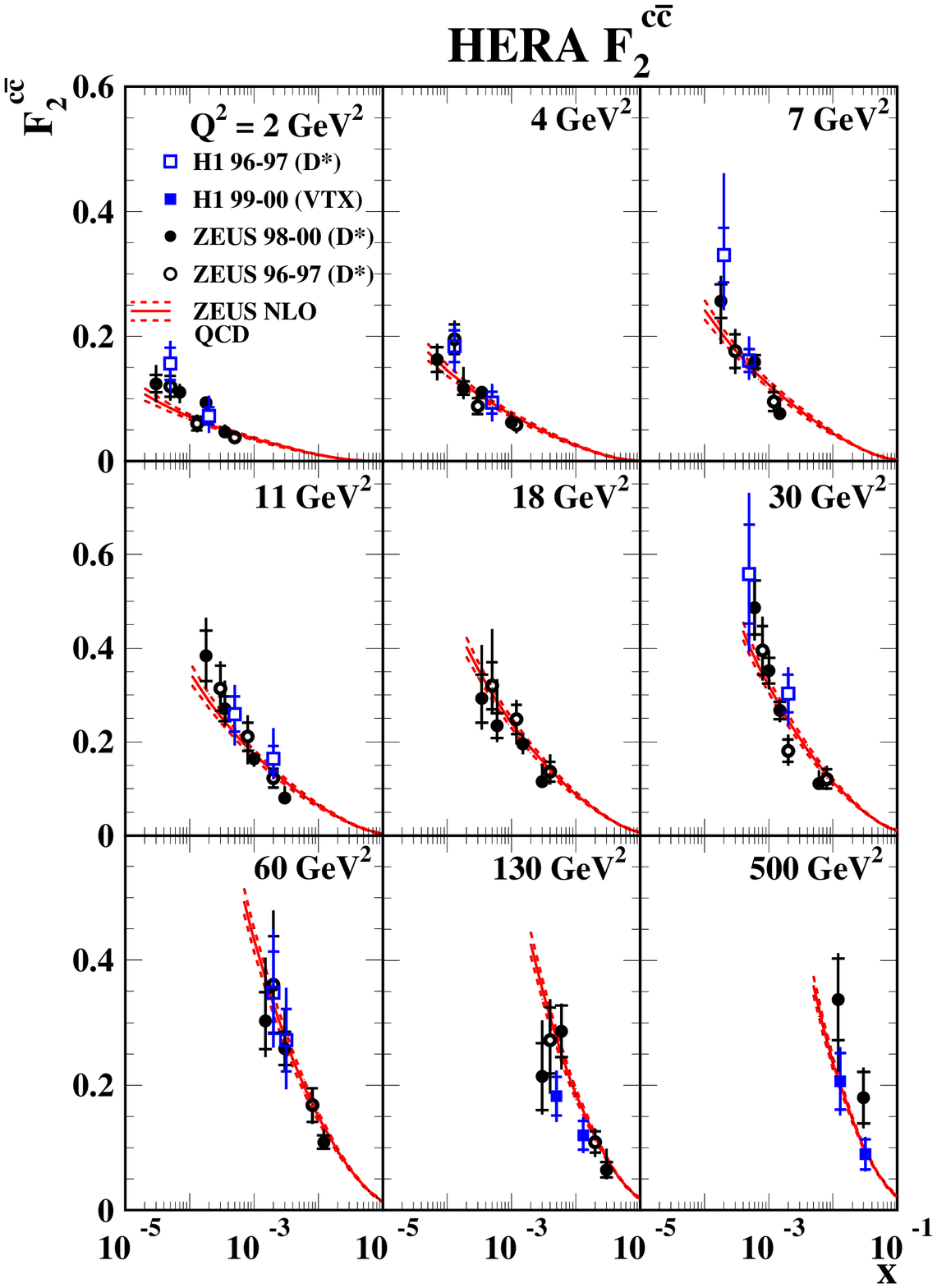, width=7.2cm,clip=true}}
\put(-0,8.2){\makebox(0,0)[l]{a)}}
\put(7.7,9.0){\makebox(0,0)[l]{b)}}
\end{picture}
\end{center}
\caption{a) Fragmentation fractions obtained at HERA and compared with $e^+e^-$ measurements. b)
Charm
\white{Figure 1:$\,\,$} contribution to the structure function $F_2$, $F_2^{c\bar{c}}$, compared with the NLO QCD fit.\hfill\label{fig:fragF2c}}
\vspace*{-0.3cm}
\end{figure}
In most measurements of charm production at HERA, $D^*$ mesons are used to identify the
presence of charm quarks, exploiting the well known mass difference method. The charm
quark cross section is then determined from the measured $D^*$ cross section by using
the fragmentation fraction $f(c\rightarrow D^*)$, as measured by other experiments.
These measurements are based on the assumption of universality of charm fragmentation.\\
In recent measurements the H1 and ZEUS experiments have determined the fragmentation fractions
$f(c\rightarrow D)$ for the various charmed hadrons ($D^+$, $D^0$, $D_s$, $D^{*+}$ and 
$\Lambda_c$ (ZEUS)) in DIS or photoproduction respectively \cite{H1Dmesons,ZEUSDmesons}.
In figure \ref{fig:fragF2c}a), the fragmentation fractions obtained in $ep$ collisions
at HERA are presented. The results are compared with measurements at $e^+e^-$ colliders 
and good agreement is observed indicating that fragmentation is universal.\\ 
In the H1 measurement the shape of the differential $D$ meson cross sections was found to be
very similar for different $D$ meson species. Also the fragmentation ratios, 
the ratio of $u$ to $d$ quarks $R_{u/d}$, the strangeness suppression factor $\gamma_s$ 
and the fraction of $D$ mesons produced as vector mesons $P_v$, obtained by ZEUS and H1 
are in good agreement with those from the LEP experiments. The uncertainties in particular
on the ZEUS measurement are competitive with those from LEP.

\section{Proton Structure}
Both H1 and ZEUS have determined the charm contribution $F_2^{c\bar{c}}$ to the structure function
$F_2$ of the proton \cite{H1F2cb,ZEUSF2c}. $F_2^{c\bar{c}}$ is shown as a function of Bjorken $x$ 
in several bins of $Q^2$ in figure \ref{fig:fragF2c}b). In the past $F_2^{c\bar{c}}$ was 
always determined by extrapolation of the measured visible $D^*$ cross section into the
full phase space.
In a recent measurement the H1 collaboration made use of the H1 silicon vertex detector
to measure the fraction of charm events from the lifetime distribution in an inclusive
event sample. In this measurement the size of the extrapolation is substantially reduced. 
The results of this measurement are indicated in figure \ref{fig:fragF2c}b) as filled squares 
in the last two $Q^2$ bins. 
Generally, good agreement between the ZEUS and H1 data and the NLO QCD prediction is obtained. 
This means that the prediction of the charm contribution to $F_2$ from scaling violations is 
consistent with the $F_2^{c\bar{c}}$ measurement.

\section{Production Mechanism}
\begin{figure}[tb]
\begin{center}
\setlength{\unitlength}{1cm}
\begin{picture}(15.,9.9)
\put(-0.5,-0.3){\epsfig{file=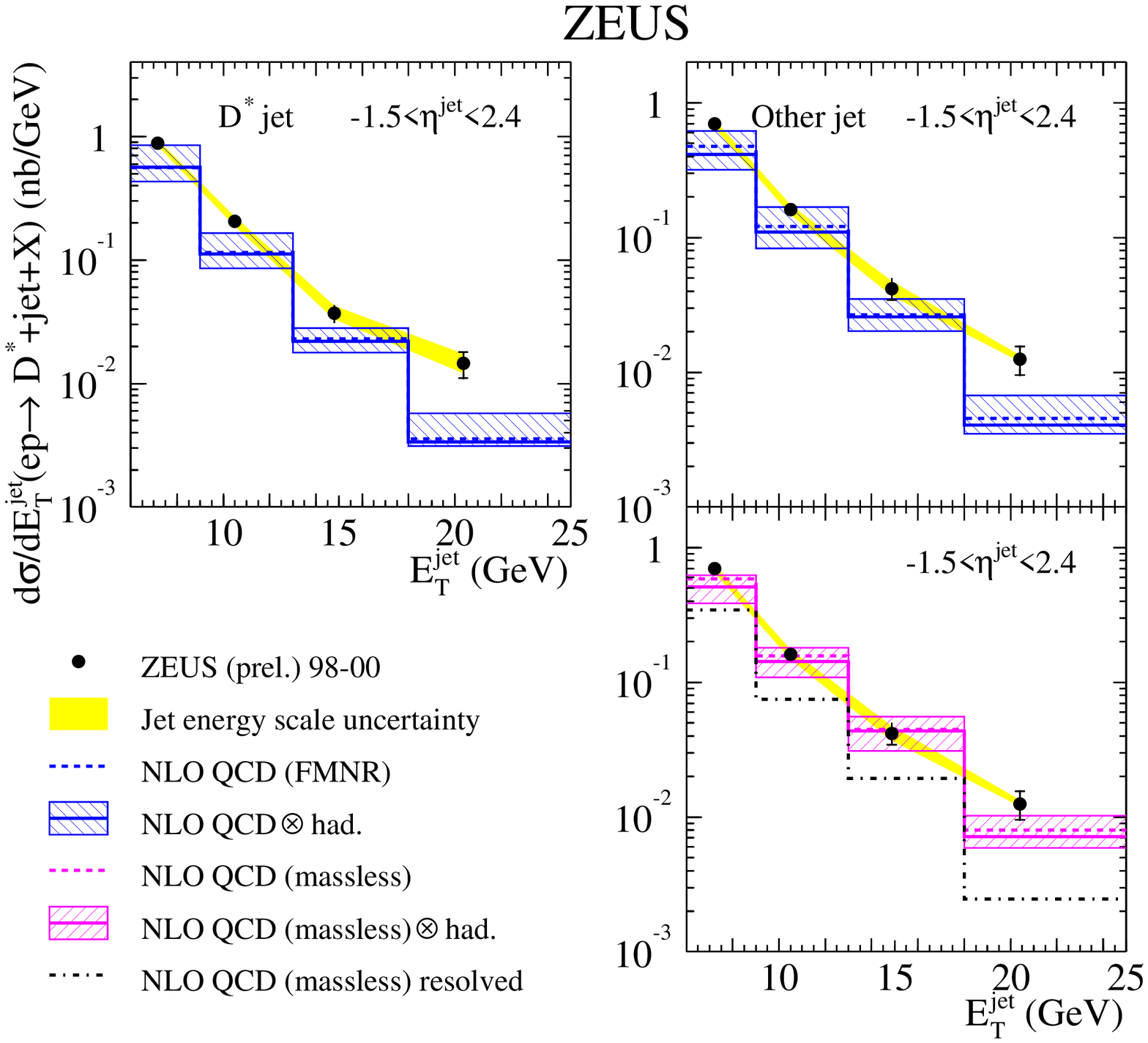, width=10cm,clip=true}}
\put(10.8,5.){\epsfig{file=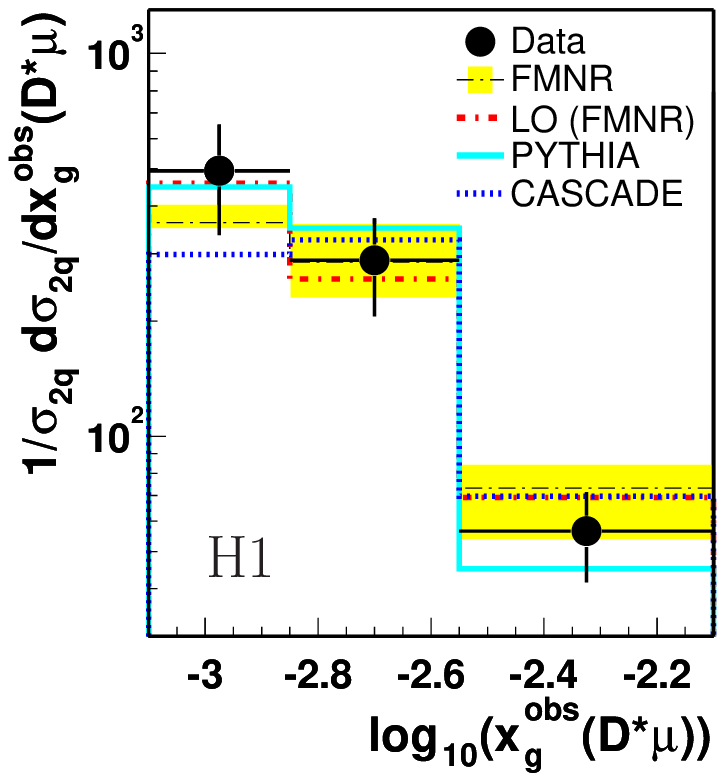, width=4.5cm,clip=true}}
\put(10.8,-0.3){\epsfig{file=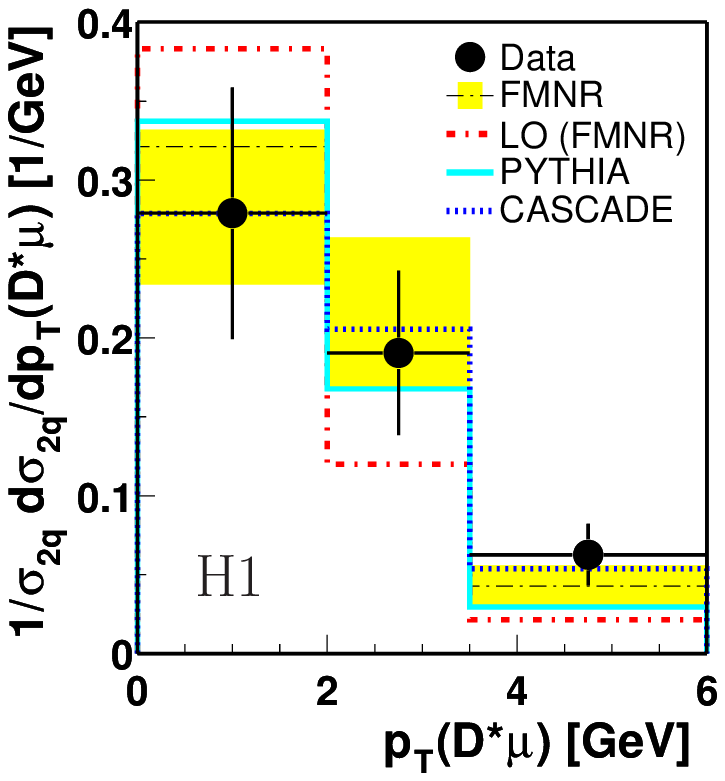, width=4.5cm,clip=true}}
\put(0.6,8.9){\makebox(0,0)[l]{a)}}
\put(10.2,9.5){\makebox(0,0)[l]{b)}}
\put(10.2,4.2){\makebox(0,0)[l]{c)}}
\end{picture}
\begin{flushleft} 
\caption{a) $D^*$ + jet cross section as a function of $E_T^{jet}$. b+c) Normalised 
$D^*$ + muon cross section as a function 
\white{Figure 2:} of $x_g^{obs}$ (b) and $p_T(D^*\mu)$ (c). The $b\bar{b}$ quark con\-ta\-mination is approximately 10\%.\hfill
\label{fig:ZEUSdstar}}
\end{flushleft}
\vspace*{-0.3cm}
\end{center}
\end{figure}
The measurement of the $D^*$ production cross section as a function of $Q^2$ for 
$1.5<Q^2<1000\,\mbox{GeV}^2$ was extended towards low $Q^2$ ($0.05<Q^2<0.7\;\mbox{GeV}^2$)
by the ZEUS collaboration using the beam pipe calorimeter \cite{ZEUSDIS}. It was found
that the NLO calculation using the ZEUS NLO fit for the 
parameterisation of the parton densities is consistent with the data over the whole $Q^2$ range. 
However, a more detailed look at the pure photoproduction data set shows some deviations 
between the data and both the massive and the matched NLO calculations \cite{ZEUSGP}. Particularly
in the intermediate transverse momentum range and in the forward direction, both theoretical predictions
are below the data.\\
If at least one jet is required in addition to the $D^*$ meson, the transverse energy of the jet,
$E_T^{jet}$, provides an additional hard scale. In an analysis of the ZEUS collaboration, carried
out in the photoproduction regime, the $D^*$ + jet cross section as a function of $E_T^{jet}$ 
was determined \cite{ZEUSET} and compared to the massive and massless NLO calculations 
(figure \ref{fig:ZEUSdstar}a)). At high jet transverse energies the data are above 
the massive calculation, while the massless NLO calculation describes the data better, 
at least in this regime.\\
In a very recent and even more exclusive analysis in the photoproduction regime,
performed by the H1 collaboration, a muon is required in addition to the $D^*$ meson \cite{H1dsmu}.
The charge and angle correlations between the $D^*$ and the muon are used to select
double tagged events. 
Due to the double tagging, details of the production mechanism can be studied.
The normalised $D^*$ + muon cross sections as a function of $x_g^{obs}$ and $p_T(D^*\mu)$
are shown in figure \ref{fig:ZEUSdstar}b+c). The quantity $x_g^{obs}$ is a good estimate of the 
relative momentum fraction of the gluon $x_g$, whereas the correlation 
is worse for $p_T(D^*\mu)$ and the transverse momentum of the gluon $k_T$.
The data are compared to the NLO calculation and the LO 
calculation (indicated as FMNR in the figure) as well as to the Monte Carlo generators
PYTHIA and CASCADE. All calculations give a reasonable description of the shape of the
$x_g^{obs}$ distribution. In the case of the $p_T(D^*\mu)$ distribution, the pure LO calculation
is too soft to describe the data, while the NLO calculation and the Monte Carlo
generators describe the data well. 

\section{Summary}
Charm production at HERA has been extensively studied. Comparisons of 
HERA data with measurements at $e^+e^-$ colliders support the assumption
that the fragmentation of charm is independent of the underlying hard physics process. 
The measurements of the charm contribution $F_2^{c\bar{c}}$ to the proton structure
are consistent with the expectations from the scaling violations of $F_2$.
NLO QCD effects are essential to predict charm photoproduction as shown
for $D^*$ + muon events. The general description is reasonable although
some details of the $D^*$ cross sections are poorly described.


\end{document}